\documentclass[notitlepage,aps,pra,twocolumn,groupaddress,10pt]{revtex4-1}

\usepackage[english]{babel}

\usepackage{graphicx}
\usepackage[colorlinks=true, allcolors=blue]{hyperref}

\usepackage{stmaryrd}
\usepackage{amssymb,amsmath,amsthm,amsfonts,amsbsy}
\usepackage{bm,bibunits,color,chngcntr,epsfig,epstopdf,graphicx,dsfont}
\usepackage{hyperref,lipsum,,makecell,mathrsfs,rotating,setspace}
\usepackage[english]{babel}
\usepackage[normalem]{ulem}

\newcommand{\be}{\begin{equation}}
\newcommand{\ee}{\end{equation}}
\newcommand{\bea}{\begin{eqnarray}}
\newcommand{\eea}{\end{eqnarray}}
\newcommand{\ket}{\rangle}
\newcommand{\bra}{\langle}

\newcommand{\I}{\mathds{1}}
\newcommand{\ra}{\rightarrow}

\newcommand{\ba}{\begin{align}}
\newcommand{\ea}{\end{align}}
\newcommand{\Tr}{\text{tr}}

\def\C#1{\mathcal #1}

\begin{document}

\title{State-adaptive quantum error correction and fault-tolerant quantum computing}
\author{Dong-Sheng Wang}\email{wds@itp.ac.cn}
\affiliation{Institute of Theoretical Physics, Chinese Academy of Sciences, Beijing 100190, China \\
School of Physical Sciences, University of Chinese Academy of Sciences, Beijing 100049, China}

\newtheorem{theorem}{Theorem}
\newtheorem{prop}[theorem]{Proposition}
\newtheorem{corollary}[theorem]{Corollary}
\newtheorem{open problem}[theorem]{Open Problem}
\newtheorem{conjecture}[theorem]{Conjecture}
\newtheorem{definition}{Definition}
\newtheorem{remark}{Remark}
\newtheorem{example}{Example}
\newtheorem{task}{Task}

\begin{abstract}
We present a theoretical framework for state-adaptive quantum error correction 
that bridges the gap between quantum computing and error correction paradigms. 
By incorporating knowledge of quantum states into the error correction process, 
we establish a new capacity regime governed by quantum mutual information 
rather than coherent information. 
This approach reveals a fundamental connection to entanglement-assisted protocols. 
We demonstrate practical applications in fault-tolerant quantum computation, 
showing how state-adaptivity enables enhanced error correction without additional measurement overhead. 
The framework provides insights into quantum channel capacities 
while offering implementation advantages for current quantum computing platforms.
\end{abstract}
\date{\today}

\maketitle

\begin{spacing}{1.0}

\section{Introduction}

The protection of quantum information constitutes a fundamental challenge in quantum information science, 
primarily due to the inevitable effects of decoherence~\cite{LB13}. 
A profound theoretical framework is required to characterize the underlying mechanisms and fundamental limits of quantum information protection.
In the classical regime, Shannon's pioneering coding theory established a foundational paradigm, 
employing entropy as a measure of information and mutual information to quantify the ultimate limit of reliable communication—the channel capacity—over noisy channels~\cite{Sha48}. 
Subsequently, diverse families of error-correcting codes have been developed, 
with certain classes asymptotically approaching this theoretical capacity bound.

Quantum versions of Shannon theory and error correction codes
have been developed these years,
and a rich spectrum of quantum channel capacities has been established~\cite{Hay17,Wil17,Wat18}.
Notably, the quantum capacity, 
specified by coherent information~\cite{SN96,Llo97,BNS98,SY08},
is the analog of classical private capacity,
both of which exhibit superadditivity, 
while the quantum mutual information 
serves as capacity measures for coding models requiring entanglement assistance~\cite{AC97,BSS+99,BSST02}. 
Compared with mutual information, the missing part of coherent information 
is the source entropy. 
What is the physical reason that 
leads to the mysterious expressions of quantum capacities? 
In this work, we study this in depth and introduce the framework of 
state-adaptive quantum error correction.

Our study is also motivated by another fact:
there is an apparent mismatch between quantum computing 
and quantum error correction; namely,
in quantum computing states and gates are often known, 
while in quantum error correction states do not need to be known. 
This operational dichotomy reflects a deeper conceptual division 
between `blind' and `visible' quantum information processing tasks,
which has been rigorously studied in contexts 
ranging from quantum source coding to computational models~\cite{Hay17,BFK09,W21_model}. 

An unknown classical bit string can be easily determined and cloned.
For a quantum state, being known or unknown makes a huge difference,
highlighted by the no-cloning theorem~\cite{WZ82,Die82}. 
The standard quantum error correction (QEC) theory requires the decoupling of environmental degree of freedom 
(e.g. eavesdropper) from the encoded information,
hence forbidding cloning and making quantum capacity be private~\cite{KL97,Dev05,HHW+08,Kle07}.
However, we realize that the decoupling and privacy are unnecessary, 
and it is possible to establish a new setting for QEC by making the encoded source 
information visible, i.e., a state-adaptive setting. 
We find that, as expected, this is the quantum analog of the classical channel capacity 
measured by mutual information.

\begin{figure}[t!]
    \centering
    \includegraphics[width=0.2\textwidth]{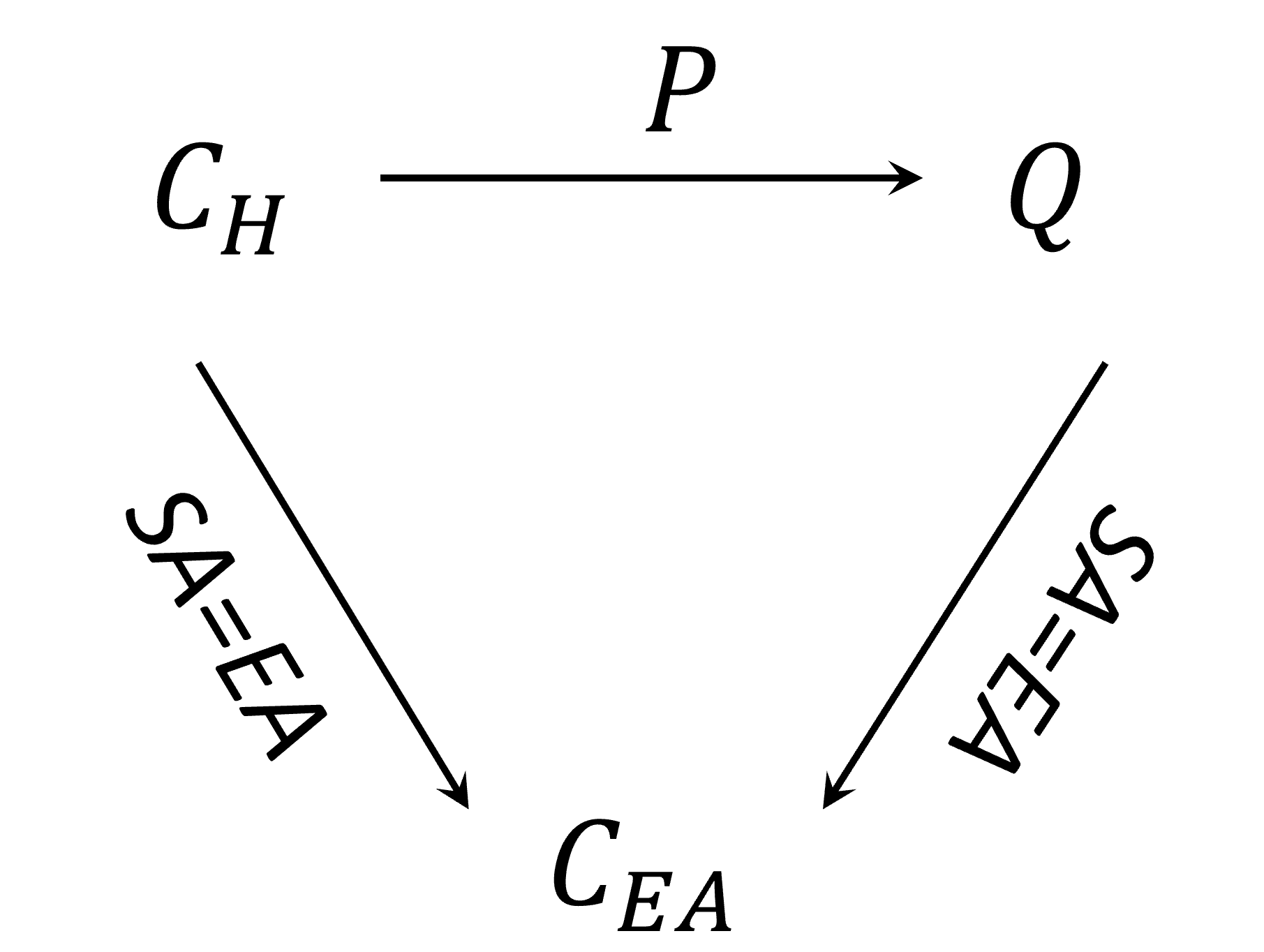}
    \caption{The diagram to show the relation among the coding models and their capacities,
with $C_H \geq P \geq Q$, $C_{EA} \geq C_H$, $Q_{EA} \geq Q$, and $C_{EA} = 2Q_{EA}$.
Here $C_H$, $P$, $Q$ are the Holevo, private, and quantum capacity of a quantum channel,
respectively.}
    \label{fig:relation}
\end{figure}

In this work, we establish the state-adaptive (SA) QEC theory.
We show that the SA quantum capacity is provided by the quantum mutual information, 
as evidenced by its equivalence to the entanglement-assisted (EA) model and a direct proof.
The equivalence highlights the role of entanglement and quantum teleportation~\cite{BBC+93}:
an unknown state can be teleported consuming entanglement,
while a known state can be directly prepared according to its classical description. 
On the dual side, the quantum dense coding~\cite{BW92} can also be substituted by SA schemes
for sending classical information over quantum channels.  
Therefore, we obtain the capacity diagram in Fig.~\ref{fig:relation}. 
By comparison, the EA model is more suitable for quantum communication scenarios 
where assisted entanglement can be pre-established, 
whereas the SA model is better suited for quantum computing contexts 
where prior knowledge of quantum states facilitates error correction.

We apply the SA model to universal fault-tolerant quantum computing based on 
stabilizer codes~\cite{Got98}. 
We propose a scheme that incorporates gate teleportation~\cite{GC99,ZLC00,DT25}, 
code switching~\cite{PR13,ADP14,Bom15,CM18}, and transversal logical gates~\cite{CCC+08,ZCC11,EK09}. 
The knowledge of logical states allows the correction of more errors 
without an obvious increase of stabilizer measurement overhead
compared to the standard scheme.
We also show that our scheme can be readily applied to current 
experimental platforms, including both photonic and solid-state ones. 

\section{Channel capacity}

Quantum decoherence and noises are described as 
quantum channels, which are 
completely positive trace-preserving maps~\cite{NC00}.
A quantum channel $\Phi$ can be represented by a set of 
Kraus operators $\{K_i\}$, or a Choi state 
$\omega_{\Phi} := \Phi \otimes \I (\omega)$
for $\omega:=|\omega\ket \bra \omega|$ as a Bell state, also known as an ebit, 
and $|\omega\ket:=\frac{1}{\sqrt{d}} \sum_i |ii\ket$, 
$d$ as the input system dimension.
It can also be represented by a unitary operator $U$ with 
$K_i=\bra i|U|0\ket$ for the index $i$ labeling an environmental basis state.
A channel $\Phi$ also has a unique complementary channel $\Phi^c$ 
by considering the process from system to the environment.

The standard quantum coding is formulated as a channel emulation task~\cite{Wat18}, 
namely, 
converting $n$ parallel uses of $\Phi$
into $k$ approximate uses of an identity channel, for positive integers $n$ and $k\leq n$.
The operation on a channel is in general a superchannel~\cite{GW07,CDP08},
which takes the form
\be \hat{\C S} (\Phi)(\rho)= \text{tr}_{a} \C V_2  (\Phi \otimes \I) \C V_1 (\rho),  
\label{eq:superchannel} \ee
for a pre- isometry $\C V_1$ that requires an ancilla $a_1$
and a post isometry $\C V_2$ that requires another ancilla $a_2$, and $a:=a_1a_2$.
A noise-free quantum memory between the pre- and post operations is in general required. 
Note we put a hat on the symbols for superchannels. 

Given a noise channel $\Phi$,
a quantum coding task can then be defined as 
a superchannel $\hat{\C S}$ so that 
\be F(\omega^{\otimes k},(\hat{\C S} (\Phi^{\otimes n})\otimes \I^{\otimes k})(\omega^{\otimes k}))
\geq 1-\epsilon, \label{eq:codeerror}\ee 
with $\epsilon\in [0,1]$ and
the state fidelity function $F(\rho,\sigma):=\|\sqrt{\rho}\sqrt{\sigma}\|_1^2$,
with $\|\cdot\|_1$ denoting the trace norm~\cite{MW14,LM15,WFD19,WLWL24}.
The fidelity above is the fidelity between ebits $\omega^{\otimes k}$
and the Choi state of $\hat{\C S} (\Phi^{\otimes n})$,
also known as the average entanglement fidelity~\cite{Sch96,Kle07,ZHLJ24}.
For any $\epsilon$,
the supremum of all achievable coding rate $\alpha:={k}/{n}$ is known as 
the quantum capacity of the channel for such a coding protocol.

The framework above includes the two standard quantum capacities as special cases:
the usual quantum coding is the factorized setting with a pre- encoding isometry
and a post decoding channel, and the capacity is 
\be Q(\Phi)= I_{c_r}(\Phi) \label{eq:capacityq}\ee 
for $I_{c_r}(\Phi):=\lim_{n\ra \infty}\frac{1}{n} \max_\rho I_c(\rho,\Phi^{\otimes n})$ 
denoting the regularized coherent information, and 
$I_c(\rho,\Phi):=S(\Phi(\rho))-S(\Phi^c(\rho))$ is the coherent information
for an input source state $\rho$ over a channel $\Phi$~\cite{SN96,Llo97,BNS98},
and the entanglement-assisted (EA) coding is the more generic setting with the capacity
\be Q_{EA}(\Phi)=\max_\rho \frac{1}{2} I(\rho,\Phi) \label{eq:capacityea}\ee 
for $I(\rho,\Phi):=S(\rho)+I_c(\rho,\Phi)$ as the quantum mutual information,
and $S(\rho)$ as the von Neumann entropy of the state $\rho$~\cite{AC97,BSS+99,BSST02}.

The coding task above allows approximate quantum error correction,
while the inaccuracy $\epsilon$ is required to be vanishing~\cite{LNC+97,SW02,BO10,NM10,WZO+20}.
When error correction is exact, 
a set of errors $\{K_i\}$ on a code satisfies the
Knill-Laflamme condition
\be PK_i^\dagger K_j P=c_{ij}P, \label{eq:kl}\ee 
for $P=VV^\dagger$ as the projector on the code space $\C C$,
and $V:\C C\ra \C H$ as the encoding isometry from the logical code space $\C C$
to the physical space $\C H$, with $\text{dim}\C C \leq \text{dim}\C H$~\cite{KL97}.
The matrix $\rho_\text{KL}:=[c_{ij}]$ 
can be understood from the complementary channel point of view 
as (the transpose of) 
an environment state $\rho_E=\sum_{ij} \text{tr}(PK_i^\dagger K_j)|j\ket\bra i|/\text{tr}P$,
while $\Tr\rho_E$ might be smaller than 1.
By diagonalizing it, this leads to $PF_i^\dagger F_j P=p_i\delta_{ij}P$, 
with effective errors $F_i$ whose action is actually unitary. 
The rank of $\rho_E$ can be smaller than the cardinality of the index $i$,
leading to the notable degeneracy phenomena for quantum codes~\cite{Got98,SS07}.
When the condition is approximately satisfied,
the state $\rho_E$ is also almost decoupled from the logical information~\cite{Dev05,HHW+08,Kle07},
and the decoder is essentially a Petz recovery channel correcting the set of typical errors~\cite{Pet86,OP93,Wil17,BK02,NM10}. 

The decoupling of $\rho_E$ makes quantum capacity automatically private.
Recall that for classical Shannon theory there are classical capacity $C$ and private capacity $P$
of a classical channel, with $C\geq P$.
Both the quantum capacity $Q$ and the private capacity $P$ show superadditivity~\cite{Dev05,SY08}, 
and degradable channels can be defined to enforce subadditivity~\cite{DS05}. 
Now we introduce the state-adaptive quantum capacity, $Q_{SA}$, 
as the analog of $C$, 
which does not require the decoupling of $\rho_E$.

\section{State-adaptive quantum error correction}

We define state-adaptive quantum coding as the task to convert 
 $n$ parallel uses of $\Phi$ into $k$ approximate uses of an identity channel so that
\be F(\hat{\C S}_\rho (\Phi^{\otimes n})(\rho), \rho)\geq 1-\epsilon,  
\label{eq:saf}\ee 
with $\forall \rho \in \C D(\C C)$, 
$\hat{\C S}_\rho$ is a coding superchannel that depends on $\rho$,
and $\text{dim}\C C=d^k$, $k\leq n$.
For simplicity, we consider the qubit case $d=2$,
but the results extend to any $d$.
This requires that the source state $\rho$ being known 
(to the encoder and decoder),
and even the environment, if acted as an eavesdropper,
can make a copy of it without the constraint of the condition~(\ref{eq:kl}).
This is in sharp contrast with the usual setting of quantum coding, 
but it does not affect the universality of quantum computing,
as we will study later on.
Compared with condition~(\ref{eq:codeerror}),
the state-adaptive setting is more of a state conversion task
rather than a channel conversion task.
It describes the fault-tolerant logical state generation task 
either for computation or bipartite communication.

Below we will prove the capacity for the state-adaptive coding model,
with a first proof based on the simulation of the EA model,
and then a direct Shannon-style proof. 
First, note that from circuit compiling theory~\cite{NC00} 
a $k$-qubit state in general requires $4^k$ parameters (modular constant),
hence $2k$ bits for a classical description of a circuit to generate it,
modular $\log\log \frac{1}{\epsilon} \in o(n)$ bits to describe each parameter 
within an accuracy $\epsilon$.
Only the $2k$ bits matter in the large-$n$ limit 
and the factor of 2 is essential. 
This is because quantum states are operators on instead of vectors in Hilbert spaces.
We will also show two crucial facts:
that feedback does not change the capacity;
and that there is no need to use EA in the state-adaptive setting, i.e.,
$\hat{\C S}_\rho$ factorizes into a pre- isometric encoding $\C V$ and a post decoding $\C D$,
so the condition~(\ref{eq:saf}) 
reduces to $F(\C D \Phi^{\otimes n}\C V(\rho), \rho)\geq 1-\epsilon$.
It is the decoder $\C D$ that depends on $\rho$, $\C V$, and $\Phi$.
This establishes the state-adaptive quantum capacity as the analog
of the classical capacity of a classical channel.

\section{Equivalence with EA setting}

\begin{theorem}
The state-adaptive quantum capacity $Q_{SA}(\Phi)$ of a quantum channel $\Phi$ is half of
the maximal quantum mutual information 
\be Q_{SA}(\Phi)=\max_\rho \frac{1}{2}I(\rho, \Phi). \ee 
\end{theorem}
\begin{proof}
We prove this by the simulation of EA classical coding task whose capacity
is $C_{EA}=2Q_{EA}=\max_\rho I(\rho, \Phi)$~\cite{AC97,BSS+99,BSST02}.
If $\alpha$ is an achievable rate in EA classical coding,
a protected message $x$ of $2k$ bits can be treated as the classical description 
of a $k$-qubit quantum state $\rho$, e.g., a quantum circuit diagram used to generate $\rho$
from a trivial state,
then the receiver can prepare $\rho$ according to $x$, 
hence completing a SA quantum coding task. 
This means $Q_{SA}\geq C_{EA}/2$.

On the other hand, if $\alpha$ is an achievable rate in SA quantum coding,
it can be combined with an outer quantum dense coding scheme~\cite{BW92} to send classical message,
since each $2k$-bits message $x$ corresponds to a known $k$-qubit state 
$\rho_x$ which serves as the source in the SA quantum coding,
hence completing an EA classical coding task. 
This means $Q_{SA}\leq C_{EA}/2$.
\end{proof}

To better understand the proof, 
note that in the relation $Q_{SA}\geq Q_{EA}$,
the left hand side cannot be replaced by $Q$, the capacity for the standard quantum coding,
wherein the source state $\rho$ is unknown
so the message $x$ cannot be treated as the classical description of it.
Also the right hand side of $Q_{SA}\leq C_{EA}/2$ cannot be replaced by 
the Holevo capacity $C_H$~\cite{Hol99},
wherein $\rho_x$ is an encoding of $x$,
and a measurement, which may lose information, is needed to extract $x$ from $\rho_x$. 

\section{Direct proof of state-adaptive quantum capacity}

To further elucidate the state-adaptive setting,
we provide a direct proof of the state-adaptive quantum capacity,
shown in the appendix which also contains 
examples for a few channels. 
Following standard approaches~\cite{Hay17,Wil17,Wat18},
the method for the decoding also uses a Petz map~\cite{Pet86,OP93,Wil17,BK02,NM10,GLM+22}.
Given a state $\rho$ and a channel $\Phi$, 
the Petz map $\C R_{\rho,\Phi}$ 
is specified by the set of Kraus operators $\{R_i\}$ with 
\be R_i=\rho^{1/2} K_i^\dagger \Phi(\rho)^{-1/2}. \ee
Special forms of it are often used in error correction, 
including the pretty-good measurement and transpose channel which uses a code projector $P$
instead of $\rho$ for its Kraus operators~\cite{BK02,NM10}.
Actually, the Petz map $\C R_{\rho,\Phi}$ can fully recover a known
input state $\rho$ with $\C R_{\rho,\Phi}\Phi(\rho)=\rho$.
This may imply that it is unnecessary to use coding to protect $\rho$ against $\Phi$.
However, implementing $\C R_{\rho,\Phi}$ is nontrivial and 
can also suffer from noises, hence cannot guarantee fault tolerance without coding. 
In the coding model, although the noise channel $\Phi$ only appears after the encoding, 
noises within the encoding and also decoding can be dealt with~\cite{Shor96,Got10}, 
rendering the whole coding scheme fault tolerant.

\section{Relation among capacities}

The knowledge of the input data source $\rho$ is powerful. 
It is easy to see there is no need to use EA in the state-adaptive model 
since given the knowledge of the input source $\rho$, encoding $V$, and also the noise channel $\Phi$,
the Petz map gives the optimal recovery~\cite{BK02,NM10}. 
For the proof, only a partial Petz recovery is needed to correct typical errors.
The role of EA is understood from quantum teleportation and its extension:
an unknown state $\rho$ can be transmitted by sending the classical bits from Bell measurements
with assisted Bell states,
while optimal assisted state can be constructed which factorizes into a sum of 
Bell states each in a typical subspace~\cite{Wil17}.
Feedback would not increase the capacity $Q_{SA}$ either for the same reason,
and indeed it has been proven that feedback cannot change $C_{EA}$~\cite{Bow04}.

As dense coding is the dual of teleportation~\cite{CL24},
one shall expect SA can also replace dense coding for sending classical information.
This is indeed the case. 
This refers to the Holevo setting, which encodes bits $x$ as states $\rho_x$,
while EA can boost the capacity to $\chi_{EA}$,
which is equivalent 
to the quantum mutual information in the large-$n$ limit~\cite{Hol02,Wat18}. 
If SA is applied, the bits $x$ is treated as the classical description of $\rho_x$,
and we have argued that there is no need of EA to send $\rho_x$,
hence SA plays the role of EA for the Holevo setting.
The EA classical capacity $C_{EA}$ can then also be viewed as SA classical capacity. 
In all, this yields the triangular relation diagram shown in Fig.~\ref{fig:relation}.
To boost the capacity for sending classical or quantum information,
the role of EA and SA is equivalent,
and the achievable classical capacity is the maximal $I(\rho_A:\rho_B)$,
with quantum capacity half of it.
Note here we denote $I(\rho_A:\rho_B)$ the same as $I(\rho,\Phi)$
in order to compare with the classical expression.

The state-adaptive model can also be used to describe the classical capacity, 
leading to a \emph{state generation} interpretation of the standard classical communication task.
Namely, to receive a set of message $\C M=\{m\}$ from Alice, 
Bob has to distinguish them without \emph{a priori} knowing each of them.
This can be re-interpreted as sending the description of a message $[m]$,
which apparently is the same as $m$ itself,
and the task of Bob is to prepare $m$ upon receiving $[m]$.
Such an interpretation is possible because
the source entropy $S_A$ not only determines the size of the input typical set, 
but can also be viewed as a measure of the average complexity to generate a state,
i.e., its average Kolmogorov complexity~\cite{CT06}.
This also builds the connection between communication and computation tasks,
hence shifting to a computational understanding of channel capacity.

\section{Transversal fault-tolerant universal quantum computing via coding switching}

We now study how to apply the state-adaptive quantum coding model 
for fault-tolerant universal quantum computing.
Indeed, in quantum computing we usually work in the visible setting:
start from a known initial state, 
apply known gates, and then perform known measurements.
On the contrary, the current quantum error-correction practice
is both state and noise blind.
Below we will show how to use the visible knowledge
to improve the error-correction performance.

The major class of quantum codes are stabilizer codes~\cite{Got98}.
A stabilizer code is a code specified by a stabilizer group.
Namely, a $[[n,k,d]]$ stabilizer code $\C C$ encoding $k$ logical qubits into $n$ physical qubits 
is defined by $n-k$ independent commuting stabilizers $\bra s_i \ket$,
which generate an Abelian subgroup of the $n$-qubit Pauli group.
The decoder is to measure its commuting stabilizers by projectors $P_i=(\I+s_i)/2$,
and the identified errors are effectively Pauli errors, 
no matter what the physical noise is since a Kraus operator can be 
expanded as a superposition of Pauli errors.
In this sense, it is neither channel-adaptive nor state-adaptive.
A gate $U$ is logical if it 
commutes with the code projector $P=\prod_i P_i$ for $[U,P]=0$, 
and for a stabilizer code,
the elementary logical gates $X_L$ and $Z_L$  
are usually transversal product of Pauli operators. 

A stabilizer state $|\psi\ket$ is also specified by a stabilizer group,
and it can be viewed as a one-dimensional code. 
To distinguish a state from a code, 
we call the stabilizers for a state as `correlators,'
a term from the early age of graph state~\cite{HDE+06}.
If a logical state from a code $\C C$ is a stabilizer state, 
then we can use this feature to improve error correction. 
Namely, we can measure the correlators $\{c_i\}$ of 
a stabilizer state $|\psi\ket \in \C C$, 
which can apparently correct more errors than measuring the stabilizers $\{s_i\}$ of code $\C C$.
For instance, Laflamme's 5-qubit code $[[5,1,3]]$~\cite{Laf96}
has 4 stabilizers of the form $ZXXZ$,
but the logical state $|0_L\ket$ is stabilized by 5 
correlators of the form $ZXZ$,
which is a 1D cluster state~\cite{RB01}.
Actually, it is established that any stabilizer state is a graph state~\cite{Sch01,VDD04},
hence we can use correlators of it for error correction. 

Operations that preserve the set of stabilizer states are known as Clifford operations,
which are not universal for quantum computing, though.
To achieve universality, other gates are needed, 
such as the $T$ gate and the Toffoli gate which belong to the 
3rd level of the Clifford hierarchy~\cite{GC99}.
The error correction of stabilizer codes can guarantee the fault-tolerance of logical Clifford gates, 
while other methods are needed for fault-tolerant $T$ gate~\cite{BK05,CSS09,PR13,WZO+20,PSB+21,KT23}. 
Here we present a scheme that combines a few primitives~\cite{GC99,PR13,ZLC00,ADP14,Bom15,CM18,DT25}, 
including gate teleportation and code switching by measurement.

\begin{figure}[t!]
    \centering
    \includegraphics[width=0.25\textwidth]{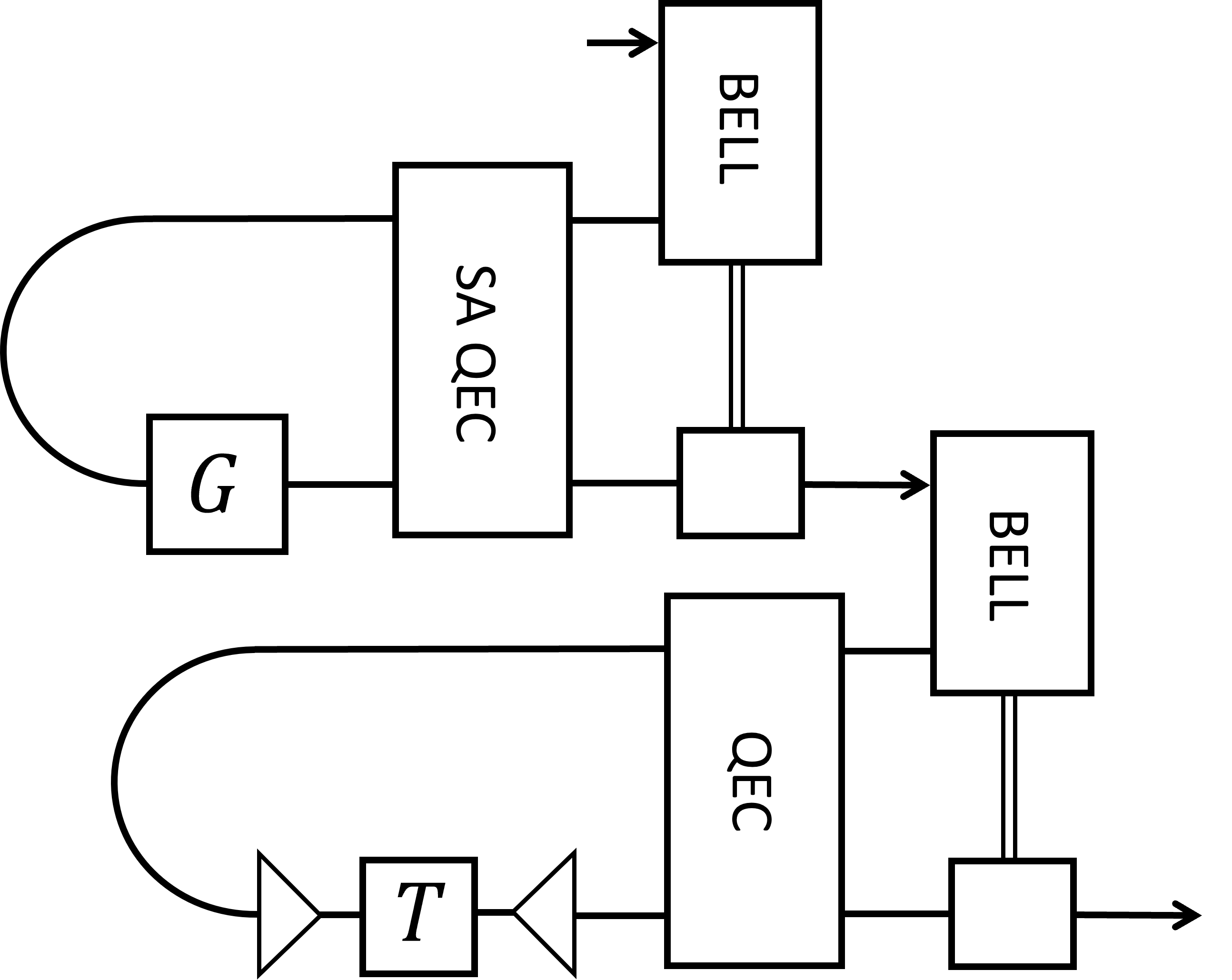}
    \caption{Schematics for the fault-tolerant quantum computing scheme 
    based on gate teleportation and code switching. 
    A transversal logical Clifford gate $G$ is stored in its Choi state $|G\ket$, 
    protected by SAQEC and executed by gate teleportation.
    A transversal logical gate $T$ is performed by a back and forth code switching (the triangles)
    and transversal $T$ on the new code, protected by usual QEC and executed by gate teleportation.
    The arrows show the time flow.
    }
    \label{fig:tvcs}
\end{figure}

The scheme is shown in Fig~\ref{fig:tvcs}. 
We use transversal logical gates by code switching between stabilizer codes,
which is well studied and efficient algorithm for switching is also developed~\cite{CM18}.
Probably the smallest universal example is the switching between Steane code
(which implements Clifford gates) and Reed-Muller 15-qubit code 
(which implements $T$ gate)~\cite{ADP14}. 
If taken together, the switched codes can be viewed as different gauge fixing of a certain 
subsystem code, which is defined by a set of non-commuting gauge operators.

Each block of logical Clifford gates $G$ is stored in its Choi state $|G\ket$ as program state
which is protected by SAQEC. 
A state $|G\ket$ is a graph state, and it is important to note that
transversal entangling gates such as CNOT only increase the weight of a correlator by a constant,
and the SAQEC is jointly applied on the entangled code blocks, 
while traditional QEC can be separable for each code block. 
On an input $|\psi\ket$, quantum teleportation by logical Bell measurement
will yield $G|\psi\ket$. 
A logical $T$ gate is performed by a back and forth code switching and transversal $T$ on the new code. 
The state is not a stabilizer state,
so standard QEC must be applied either for the old or the new code. 

The reason to use teleportation for Clifford gates is that 
once a $T$ gate is applied, the state is not a stabilizer state anymore,
and it becomes rather infeasible to use the knowledge of the state
and apply the SAQEC directly. 
Instead, SAQEC can be applied on $|G\ket$ to induce a clean $G$ on an input $\rho$.
Note that in order to avoid $\rho$ itself decohering fast,
the state $|G\ket$ needs to be prepared offline and maintained clean.
To realize a circuit, 
it appears the number of qubits scales with the spacetime cost of the circuit 
due to teleportation. 
If the measured qubits in teleportation can be refreshed 
and reused to generate program states,
this can reduce the space cost.

\section{Conclusion}

In all, we proposed a framework of state-adaptive 
quantum error correction (SAQEC) 
which is a quantum analog of classical error correction.
We showed that standard stabilizer codes can be employed in the setting of SAQEC,
which can achieve higher coding rate and error threshold, 
without a significant overhead. 
By definition, an error threshold is the value of $\lambda$ at which 
a capacity of a parametrized channel $\Phi(\lambda)$ becomes zero.
It is clear to see SAQEC allows an error threshold higher than that of the standard
QEC.

There are also two important issues to explore further. 
First, as we mentioned QEC based on stabilizer codes is not channel-adaptive. 
As stabilizer measurements detect Pauli errors and effectively twirl a noise channel, 
this reduces its capacity. 
Therefore, exploring channel-adaptive code design remains a promising direction to improve QEC. 
Second, not only the decoder $\C D$ can depend on the encoder $V$, channel $\Phi$,
and source state $\rho$, the encoder $V$ can also depend on both $\Phi$ and $\rho$.
We expect that there are good quantum codes with low overhead 
that are both channel-adaptive and state-adaptive.

Our finding is rather surprising. 
It resolves a longstanding question in the field: 
why quantum mutual information serves as the capacity measure for the 
entanglement-assisted coding model, 
which does not have the classical analog. 
Our result emphasizes the importance of pre-knowledge of quantum states and operations, 
and may also benefit the study of the superadditivity phenomena in Shannon theory.

\section{Acknowledgement}
This work has been funded by
the National Natural Science Foundation of China under Grants
12447101 and 12105343.
Many thanks to R. Laflamme for his insights on quantum error correction.

\end{spacing}

\appendix 

\section*{Appendix}
\label{sec:app}

\subsection{Channel capacity}

The state-adaptive (SA) model fits into the rich family of quantum channel coding models.
This is shown in Table~\ref{tab:my_label}.
The state-adaptive quantum capacity is the analog of classical capacity, 
both of which are not private, 
and the usual quantum capacity is the analog of private capacity,
and the one in the middle is the Holevo model.

In the classical setting the encoding is from bits to bits, with a source entropy $H(X)$,
similarly for the state-adaptive setting which has a source entropy $S(\rho)$.
In the Holevo setting the encoding is from bits to qubits, still 
with a source entropy $H(X)$,
but each bit string $x$ is mapped to a state $\rho_x$ with $x$ as its \emph{label},
and measurement is required to recover $x$ from $\rho_x$.
The Holevo capacity is the mutual information of a classical-quantum state,
which is not subadditive in general, 
while does not require a fully decoupling of the environment~\cite{DST14}.
The quantum capacity, and also the private capacity of either a classical or a quantum channel,
all require the decoupling of the environment state.

The analog with classical case even extends to nontrivial computation beyond merely error correction.
Namely, the quantum code switching scheme also has a classical analog.
For a classical code $V: m \mapsto b_m$ that encodes each $k$-bit message $m\in \C M$ 
as a $n$-bit codeword $b_m$,  
a computation with it enables permutations among them.
This can be viewed as a code switching by assigning an encoding $V_m$ to each $m$,
and the computation involves switching over the set $\{V_m\}$.
Each $V_m$ could also depend on $m$.
Therefore, for the quantum case 
switching among a few codes can also be used,
with each code supporting one or a few logical gates with good features,
such as transversality.
In particular, 
non-additive codes with transversal logical gates can also be used~\cite{KT23}.
The net performance would depend on the cost of coding, switching, 
error threshold, etc. 

There is also a key difference between
the classical setting and the quantum SA setting for the role of the `environment' $E$.
Namely, $E$ can make a copy of a bit without disturbing it,
but if it tries to make a copy of a source state $\rho$, 
which is known to $A$ and $B$ but not $E$,
it has to rely on its state $\rho_E=\sum_{ij}\text{tr}(\rho K_i^\dagger K_j)|j\ket \bra i|$.
If $E$ knows each $K_i$ and $\{K_i^\dagger K_j\}$ forms an operator basis,
then $E$ can reconstruct $\rho$ by measuring the values $\text{tr}(\rho K_i^\dagger K_j)$,
which is to perform a state tomography of $\rho_E$.
This requires a large sampling cost.
That is, there is a \emph{sampling barrier} for $E$ to make a copy of source state 
even if $E$ is not decoupled from the source system.

\begin{table}[]
\vspace{0.0cm}
    \centering\footnotesize
    \begin{tabular}{c|c|c|c}\hline
                &  Measure & Additivity & Decoupling   \\ \hline
    Classical     &  $I(X:Y)$ & Sub  &  No \\ \hline
    SA quantum  &  $I(\rho_A:\rho_B)/2$     &  Sub   & No \\ \hline
    Holevo   & $I(X:\rho_B)$        &  Super     & No \\  \hline
    Private & $I(X:Y)-I(X:E)$       &  Super     &  Yes \\  \hline
    Quantum & $I(\rho_A:\rho_B)-I(\rho_A:\rho_E)$        &  Super     &   Yes \\  \hline
    \end{tabular}
    \caption{Channel capacities with key features. 
    It shows the (SA) quantum capacity of a quantum channel 
    as the analog of (classical) private capacity of a classical channel,
    while the Holevo setting is a hybrid model.\vspace{0cm}
    }
    \label{tab:my_label}
\end{table}

Below is a direct proof of the state-adaptive quantum capacity.

\begin{proof}
(Direct part) We need to show $\max_\rho I(\rho, \Phi)/2$ is a lower bound. 
The method is to use the packing lemma~\cite{Wil17}.
For the SA setting, 
the QEC condition is not satisfied, 
and the coding structure is a hybrid of the classical and the Holevo setting.
More precisely, let the input source be an arbitrary but known $k$-qubit state $\sigma$,
and denote its $2k$-bit classical description as $[\sigma]$.
For the i.i.d. case, $\sigma$ is randomly encoded into a product state $\rho^{\otimes n}$,
which can be viewed as the inverse of Schumacher compression~\cite{Sch95}. 
The size of the input typical set is determined by the source entropy $S_A=S(\rho)$,
and similarly for the receiver $B$.
The decoder at the output 
is a global Petz recovery channel, namely, a pretty-good measurement, 
only applied for typical sequences.
As $[\sigma]$ is perfectly correlated with the state $\rho$ itself,
the disturbance of each typical codeword by the noise channel 
is determined by the conditional entropy $S_{B|A}=S_{AB}-S_A$.
It is then clear from the packing lemma 
any rate $I(\rho, \Phi)=S_B-S_{B|A}$ is achievable 
to send the classical description $[\sigma]$,
hence $I(\rho, \Phi)/2$ for the state itself. 
For the generic case, $\sigma$ is encoded into an entangled state, 
and the size of the typical set gets smaller.
Due to the subadditivity of quantum mutual information~\cite{AC97}, 
the claimed lower bound also holds.

(Converse part) We need to show $\max_\rho I(\rho, \Phi)/2$ is an upper bound. 
The method is to consider a randomness distribution task, $D$, the capacity of which 
is an upper bound of $Q_{SA}$ since the latter can be used to simulate the former.
The task $D$ is to distribute $2k$ maximally correlated bits, as in the Holevo setting, 
but here each bit string $x$ is perfectly correlated with an input state $\rho_x$ to the channel $\Phi^{\otimes n}$,
serving as its classical description. 
It is then clear to obtain 
$I(\rho, \Phi)\geq \frac{2k}{n} - \delta$ for vanishing parameter $\delta$ in the large-$n$ limit. 
This completes the proof. 
\end{proof}

\subsection{Examples of noise channels}

Besides Pauli channels which are common in the study of quantum error correction, 
there are also a few other examples that are practically relevant~\cite{Wil17}.  
We focus on the qubit case. 
The erasure channel $\C E$ erases a qubit with error rate $p$ replacing it
with a state $\rho_e$ that is outside the state space
\be \C E (\rho) =(1-p)\rho + p \rho_e.\ee 
A notable fact in QEC is that correcting $t$ errors is equivalent to 
correcting $2t$ erasures as their locations can be identified. 
Its quantum capacity is  $Q(\C E)=1-2p$,
and SA quantum capacity is  $Q_{SA}(\C E)=1-p$.

The amplitude-damping (AD) channel $\C A$ is specified by two Kraus operators 
$A_0= |0\ket \bra 0| +\sqrt{1-\gamma}|1\ket \bra 1|$,
$A_1=\sqrt{\gamma} |0\ket \bra 1|$. 
It describes the decay from an excited state $|1\ket$ to the ground state $|0\ket$,
and can be viewed as a `partial' erasure channel. 
Actually, when encoded into two qubits as degenerate states $|0\ket\equiv |01\ket$,
$|1\ket\equiv |10\ket$,
the AD channels on the two qubits will act as an erasure channel with error rate $\gamma$.
Code design for the AD channel can then be converted to that for erasure.
Its quantum capacity is  $Q(\C A)=\max_a h(a(1-\gamma))-h(a\gamma)$,
and SA quantum capacity is  $Q_{SA}(\C A)=\frac{1}{2}\max_a h(a)+h(a(1-\gamma))-h(a\gamma)$,
for $a\in[0,1]$, $h(x)$ as the binary entropy function. 
The threshold value for $Q(\C A)$ is $\gamma=\frac{1}{2}$,
while $Q_{SA}(\C A)$ is already nonnegative for any $\gamma$.

The qubit Pauli channel $\C P$ is a random mixture of Pauli operators $\sigma_i=\{X,Y,Z\}$
and identity $\I$ with probabilities $p_x$, $p_y$, $p_z$, and $p_0=1-p_x-p_y-p_z$.
When $p_x=p_y=p_z$, it reduces to the depolarizing channel $\C D$,
and when $p_x=p_y=0$, it reduces to the dephasing channel. 
Among them, the quantum capacities for $\C P$ ad $\C D$ 
do not have a closed form due to the superadditivity of coherent information. 
Here we discuss the realization of SA capacity.

Finding capacity-approaching codes is highly nontrivial even when 
a channel capacity is computable. 
Here we study code design for the erasure channel due to its simplicity
and its `universal' property, namely, 
other channels can be converted to erasure channels with a suitable encoding.
With the three-qubit repetition code $|0\ket\equiv |000\ket$, 
$|1\ket\equiv |111\ket$, measuring the two stabilizers $Z_1Z_2$, $Z_2Z_3$
can define an erasure event: any nontrivial syndrome is treated as an effective erasure.
For $\C D$ with an error rate $p$, the effective erasure channel 
has an error rate $1-(1-p)^3\geq p$.
This is the analog of the two-qubit encoding for the AD channel above. 

For an erasure channel $\C E$ with error rate $p$,
the quantum capacity $Q(\C E)=1-2p$ as otherwise 
it would violate the no-cloning theorem~\cite{BDS97}.
In the EA model, it is increased to $Q_{EA}(\C E)=1-p$,
which means even when $p>\frac{1}{2}$, the assisted entanglement
will help for the transmission of qubits. 
For the SA model, we shall use the knowledge of the input and the channel. 
Note that $\C E$ with $p$, denoted by $\C E_p$, can be simulated by 
\be \C E_p = \frac{1}{2} (\C E_{2p-1} + \C E_{1})\ee 
for $p> \frac{1}{2}$.
Let $p_1=2p-1$, then $\C E_{p_1}$ can be simulated again via 
$\C E_{p_2}$ for $p_2=2p_1-1$ and $\C E_{1}$.
This concatenation carries on and yields $p_r=1-2^r(1-p)$.
Requiring $p_r\in [0,\frac{1}{2}]$,
this implies $p\in [1-2^{-r},1-2^{-r-1}]$ approaches the maximal value 1. 

This leads to an SA coding scheme:
with $\C E_p$ ($p\in [\frac{1}{2},1]$) simulated by the concatenation at level $r$, 
a code designed for $\C E_{p_r}$ ($p_r\in [0,\frac{1}{2}]$)
is used as an outer code, and a $r$-level concatenated repetition code is used as an inner code,
so that the source is input to $n$ parallel uses of the channels $\C E_{p_r}$,
while other channels $\C E_{1}$ only output junk data. 
It is easy to see the coding rate approaches $1-p$, 
exponentially fast with $r$.
In all, this forms an SA coding scheme simulated 
by the usual quantum coding scheme for $\C E_p$ ($p\in [0,\frac{1}{2}]$), 
with a structured input source and channel, 
treating the `bad' channels $\C E_{1}$ as free resources. 
Note that the division of channels into `good' and `bad' ones 
echoes the method of polar codes~\cite{RL09}. 

There are very noisy channels that cannot be converted into erasure channels, though. 
The class of entanglement-breaking (EB) channels are such examples~\cite{HSR03}. 
An EB channel $\C B$ is specified by a positive operator-valued measure
(POVM) $\{M_i\}$ and a set of states $\{\rho_i\}$ as 
\be \C B(\rho)=\sum_i \text{tr}(M_i\rho) \rho_i, \ee 
and its quantum capacity is zero, $Q(\C B)=0$, 
while $Q_{EA}(\C B)=C_H(\C B)=\chi(\C B)$, the optimal Holevo mutual information~\cite{Wil17}.
It is a one-shot form since entanglement does not help for the coding.
For the SA model, a source $k$-qubit state $\rho$ could be entangled, 
which will be destroyed by EB channels. 
This implies that one has to send the classical description $[\rho]$ over the EB channels, 
which then induces $Q_{SA}(\C B)$ being $\chi(\C B)$. 
There exists no SA coding that can directly protect quantum state $\rho$ over EB channels.
In other words, when $Q$ is exactly zero for a channel, 
it cannot be boosted to achieve $Q_{SA}$ in the SA setting. 
On the contrary, in the EA setting which is more suitable for communication,
the boosted capacity is due to the assisting entanglement that needs to be pre-established
and maintained noiseless. 

\subsection{Physical implementation}

Here we discuss platforms for physical implementation.
There are notable differences between solid-state platforms and the photonic platform.
It is relatively easier for solid-state platforms,
but for photonic platform with qubits encoded in discrete variable,
the CNOT gate is probabilistic~\cite{PJF01,BR05}. 
A scalable approach to resolve this is the fusion-based scheme~\cite{BBB21},
which employs fusion measurements to grow a graph state efficiently.
The graph state carries the information of the circuit to be executed, 
the encoding and code switching.
One direction through the graph is the simulated time of the circuit,
and the part of the graph at a later time can be generated gradually.
The correlators of the graph can be used for error correction,
which can correct more errors than using stabilizers of a code.

The available operations are single photon measurements
and two-photon fusion measurements.
The fusion measurements, which are 1-bit and 2-bit teleportation~\cite{ZLC00},
are used to enable conversion and measurement of stabilizers,
hence realizing code switching and error correction.
Single qubit measurements are used for final readout,
and also to enable transversal logical gates,
with Clifford gates being induced by Pauli-basis measurements
and $T$ gate by rotated projective measurements.

The usage of graph states can also be applied to solid-state platforms 
since any stabilizer state can be expressed as a graph state~\cite{RB01,Sch01,VDD04}.
A notable difference between solid-state and photonic platforms is that 
for the former, separate code blocks based on graph states 
can be used and connected by deterministic teleportation,
and for the latter, an entire logical graph state, 
containing logical CNOT gates between code blocks
and generated gradually over time, is needed.

\bibliography{ext}{}
\bibliographystyle{elsarticle-num}

\end{document}